# Evidence for a Gapped Spin-Liquid Ground State in a Kagome Heisenberg Antiferromagnet

Mingxuan Fu[1], Takashi Imai[1,2,*], Tian-heng Han[3,4], Young S. Lee[5,6]

[1]Department of Physics and Astronomy, McMaster University, Hamilton, ON L8S4M1, Canada.
[2]Canadian Institute for Advanced Research, Toronto, ON M5G1Z8, Canada.
[3]James Franck Institute and Department of Physics, University of Chicago, Chicago, IL 60637.
[4]Materials Science Division, Argonne National Laboratory, Argonne, IL 60439.
[5]Department of Physics, Massachusetts Institute of Technology, Cambridge, MA 02139.
[6]Department of Applied Physics and Department of Photon Science, Stanford University and SLAC National Accelerator Laboratory, Stanford, CA 94305.

*Correspondence to:  imai@mcmaster.ca

The kagome Heisenberg antiferromagnet is a leading candidate in the search for a spin system with a quantum spin liquid ground state. The nature of its ground state remains a matter of great debate. We conducted $^{17}$O single crystal NMR measurements of the spin $S=1/2$ kagome lattice in herbertsmithite ZnCu$_3$(OH)$_6$Cl$_2$, which is known to exhibit a spinon continuum in the spin excitation spectrum. We demonstrate that the intrinsic local spin susceptibility $\chi_{kagome}$ deduced from the $^{17}$O NMR frequency shift asymptotes to zero below temperature $T \sim 0.03J$, where $J \sim 200\,\text{K}$ is the Cu-Cu super-exchange interaction. Combined with the magnetic field dependence of $\chi_{kagome}$ we observed at low temperatures, these results imply that the kagome Heisenberg antiferromagnet has a spin liquid ground state with a finite gap.

The realization and characterization of a kagome Heisenberg antiferromagnet (KHA) with a corner-shared triangle structure (Fig.1A) is crucial for the search for a quantum spin liquid ground state (*1,2*). Spin-liquids consist of entangled pairs of spin-singlets, and do not undergo a magnetic phase transition. The successful synthesis of the structurally ideal kagome lattice of $Cu^{2+}$ ions (spin $S=1/2$) in herbertsmithite $ZnCu_3(OH)_6Cl_2$ (Fig.1B-E) (*3*) was welcomed as a major milestone (*4*). $ZnCu_3(OH)_6Cl_2$ remains paramagnetic at least down to ~50 mK (*5,6*). Moreover, inelastic neutron scattering measurements (*7*) on single crystals (*8*) demonstrated that the spin excitation spectrum does not exhibit conventional magnons, but rather a spinon continuum. Despite the recent progress, fundamental issues regarding the nature of the ground state of the KHA remain to be understood. For example, the central question on the existence of a gap in the spin excitation spectrum has not been settled. This information is critical to distinguish between the leading theories for the ground state of the $S=1/2$ KHA: a gapped spin liquid, gapless spin liquid, or valence bond solid (*1,2,9-14*)

In $ZnCu_3(OH)_6Cl_2$, weakly interacting $Cu^{2+}$ defects occupy the non-magnetic $Zn^{2+}$ sites between the kagome-layers with ~15% probability (*15*). Their contributions dominate bulk-averaged thermodynamic properties at low temperatures (*3,5,8,16,17*), making it difficult to measure the intrinsic low energy properties of this material. Similarly, the $Cu^{2+}$ impurity moments can contribute to the inelastic neutron scattering, obscuring the response of the intrinsic kagome spins at low energies (< 2meV) (*7*). NMR is an ideal local probe to investigate the intrinsic magnetic behavior under the presence of magnetic defects, as demonstrated by successful investigations of Kondo-oscillations and analogous phenomena in metals (*18*), high-temperature superconductors (*19*), and low-dimensional spin systems (*20*). Our primary goal here is to uncover the intrinsic behavior of the spin susceptibility, $\chi_{kagome}$, separately from the defect-induced local spin susceptibility, $\chi_{defect}$, using $^{17}$O NMR measurements of an isotope-enriched ZnCu$_3$(OH)$_6$Cl$_2$ single crystal. We demonstrate that the spin excitation spectrum exhibits a finite gap $\Delta = 0.03J \sim 0.07J$ between a $S = 0$ spin-liquid ground state and the excited states, where $J \sim 200$ K represents the Cu-Cu super-exchange interaction (*5,21*).

A major advantage of using a single crystal for NMR is that we can achieve high resolution by applying an external magnetic field **B**$_{ext}$ along specific crystallographic directions. In Fig. 2A we present the $^{17}$O (nuclear spin $I = 5/2$, gyromagnetic ratio $\gamma_n/2\pi = 5.772$ MHz/T) NMR lineshape measured at 295 K in $B_{ext}$ = 9 T applied along the c-axis; the temperature dependence of the lineshape is presented in Fig.2C. Unlike previously measured powder-averaged $^{35}$Cl and $^{17}$O NMR lineshapes (*22,23*), we can clearly resolve the five $I_z = m$ to $m+1$ transitions ($m = -5/2$, -3/2, -1/2, +1/2, +3/2) separated by a nuclear quadrupole frequency $\nu_Q^{(c)}$. In addition, the central peak frequency $f$ for the $I_z = -1/2$ to $+1/2$ transition is shifted from the bare resonance frequency $f_o = (\gamma_n/2\pi)B_{ext}$ by the effects of the hyperfine magnetic fields from nearby Cu$^{2+}$ sites, and the shift of the peak (marked as Main in Fig. 2A) is proportional to $\chi_{kagome}$. Such a NMR frequency shift may be expressed in terms of the Knight shift, $^{17}K^{(c)} = f/f_o - 1 = A_{hf}\chi_{kagome}$, where $A_{hf}$ is the hyperfine coupling constant between the $^{17}$O nuclear spin and the Cu$^{2+}$ electron spins. We can fit the lineshape in Fig. 2A with three sets of five peaks with three distinct values of $^{17}K^{(c)}$ and $\nu_Q^{(c)}$. That is, the presence of the Cu$^{2+}$ defects at the Zn$^{2+}$ sites results in three distinct $^{17}$O sites in ZnCu$_3$(OH)$_6$Cl$_2$. Taking into account the difference in the transverse relaxation that affects the apparent signal intensities (Fig. S1), we estimated the population of the three sites as 13(4)%, 28(5)%, and 59(8)%, in agreement with earlier $^2$D NMR observation of three corresponding sites in a deuterated $ZnCu_3(OD)_6Cl_2$ single crystal (*24*).

It is straightforward to assign these $^{17}$O signals to the nearest neighbor (NN) of the $Cu^{2+}$ defects, the next-NN (NNN) of the $Cu^{2+}$ defects, and the intrinsic Main sites located far from defects, respectively, for the following reasons. First, we recall that the $Cu^{2+}$ defects occupying the $Zn^{2+}$ sites with ~15% probability (*15*) is the cause of a large Curie-Weiss contribution $\chi_{defect}$ in the bulk magnetic susceptibility data at low temperatures (*24*). This means that ~15% of $^{17}$O (and $^2$D) at the NN sites experience an additional, transferred-hyperfine coupling $A_{hf}^{defect}$ with these $Cu^{2+}$ defect spins polarized by $B_{ext}$. As shown in Fig. 2B, the large Curie-Weiss contribution $A_{hf}^{defect} \chi_{defect}$ to $^{17}K^{(c)}$ at the NN $^{17}$O sites overwhelms the intrinsic contribution $A_{hf} \chi_{kagome}$ at low temperatures, similar to the defect induced Curie-Weiss behavior of $^2K_{NN}^{(c)}$ at the NN $^2$D sites reported earlier (*24*). Both $^{17}K^{(c)}$ and $^2K_{NN}^{(c)}$ are negative at low temperatures simply because $A_{hf}^{defect} < 0$. Second, since the abundance of the NNN $^{17}$O sites should be twice of the NN as shown in Fig.1C and 1E, we can naturally assign the 28(5)% signal to the NNN. The remaining 59(8)% arises from the intrinsic Main sites. We emphasize that our high resolution single crystal NMR data show no evidence for the presence of $Zn^{2+}$ anti-site defects occupying the $Cu^{2+}$ sites within the kagome-plane, in disagreement with earlier powder-averaged $^{17}$O NMR data, which were obtained from lower-resolution measurements (*23*). The previous powder work (*23*) misidentified the NN peaks as arising from singlet "D-sites" induced by anti-site defects in the kagome plane, as explained in detail in (*25*).

The NMR properties at the NNN and main sites are very similar, and we could clearly distinguish them only near 295 K where the NMR lines are narrow. Their similarity suggests that the spin polarization induced by the $Cu^{2+}$ defects decays very quickly within the kagome plane, unlike the long-range nature of the Kondo oscillations in metals (*18*). Our observation is consistent with the expectation that the spin-spin correlation length is comparable to the Cu-Cu distance in the KHA (*7,26,27*). The primary effects of the quenched randomness of the $Cu^{2+}$ defects on the Main peaks are therefore the magnetic line broadening in proportion to $\chi_{defect}$ (Fig.S2). The broad high frequency tail of the Main peak observed below 4.2 K in 9 T (Fig.2C, Fig.S3-S4), however, may also indicate that the $Cu^{2+}$ defects induce a short-range spin density oscillation within the kagome plane, and some NNN sites have a large positive $^{17}K$ (Fig. S4).

In the field geometry of $\mathbf{B}_{ext} \parallel c$, accidental superposition of the NN peaks prevented us from resolving the central part of the Main peak below ~70 K (Fig.2C). Instead, we found that $\mathbf{B}_{ext} \parallel a^*$ is the ideal geometry for investigating $\chi_{kagome}$ at low temperatures. For the $\mathbf{B}_{ext} \parallel a^*$ geometry(Fig.1B), the direction of $\mathbf{B}_{ext}$ is at a 60° angle to the Cu-O-Cu bond of two $^{17}O$ sites ("Main1") within each triangle, and is orthogonal to the bond of the remaining $^{17}O$ site ("Main2"). As a consequence (Fig.3A), the $^{17}O$ NMR signals from the Main1 and Main2 sites appear separately with the integrated intensity ratio of 2:1 and different magnitudes of $^{17}K^{(a^*)}$ and the quadrupole splitting ($\nu_Q^{(a^*)}$~ 8 kHz for Main1 and $\nu_Q^{(a^*)}$~ 520 kHz for Main2). Normally, the partitioning of the $^{17}O$ NMR lines into Main1 and Main2 would merely complicate the overall spectrum. Here, thanks to the small value of $\nu_Q^{(a^*)}$~ 8 kHz, all five of the $I_z = m$ to $m+1$ transitions of Main1 are bundled into a large peak that we could resolve down to 1.8 K (*~0.01J*). Moreover, we were able to selectively measure the Main1 lineshape (Fig. 3B) by exciting all five transitions at once with longer radio-frequency pulses (*25*); the isolated Main1 peak frequency in Fig.3B agreed well with that of Fig.3A, providing extra confidence in our line assignment.

In Fig.4A, we summarize the temperature dependence of $^{17}K^{(a*)}$ of the Main1 sites measured in $\mathbf{B}_{ext} = 3.2$ T $\parallel$ a*. The results of $^{17}K^{(a*)}$ measured in $\mathbf{B}_{ext} = 9$ T $\parallel$ a* (Fig. 2B) are nearly identical, except below ~10 K as discussed below. The superposition of the large Main1 peak made accurate measurements of $^{17}K^{(a*)}$ at the central transition of Main2 difficult, but the frequency shift of the isolated uppermost satellite peak of Main2 shows a similar trend with temperature (Fig. 3A and Fig. S5). Unlike the bulk averaged magnetic susceptibility data (*3,5,8*), $^{17}K^{(a*)}$ decreases below ~60 K. Moreover, $^{17}K^{(a*)}$, and hence $\chi_{kagome}$, asymptotes to zero below ~5 K (*~0.03J*), implying the absence of thermally accessible excited states. The results of the $^{17}O$ nuclear spin-lattice relaxation rate $1/T_1$ are also consistent with diminishing intrinsic spin excitations toward $T = 0$ (Fig. S6). Combined with the fact that the structural analysis at low temperatures based on Rietveld refinement revealed no hint of the formation of a large unit cell expected for valence bond solids (*15*), we conclude that $ZnCu_3(OH)_6Cl_2$ has a quantum spin liquid ground state with the total spin $S = 0$, and its excitation spectrum has a finite gap.

To estimate the magnitude of the gap $\Delta$, we fitted $^{17}K^{(a*)}$ below 4.2 K to an exponential function with a pre-factor $T$, $^{17}K^{(a*)} \sim T \cdot \exp(-\Delta/T)$, as shown in Fig.4B; we obtained $\Delta \sim 6.8$ K for $B_{ext} = 3.2$ T. The constant background term $^{17}K_{chem}$ arising from the chemical shift is generally negligibly small ($\sim \pm 0.02$ %) at the $^{17}O$ sites in copper-oxides (*28*), and does not affect our estimation of $\Delta$ significantly. We included the pre-factor $T$ to phenomenologically account for the effects of antiferromagnetic spin correlations that tend to suppress the spin susceptibility even without a gap. The same pre-factor $T$ also arises for the uniform Pauli spin susceptibility of spinon excitations in the gapped Dirac Fermion model (*11*). It is important to realize that the magnetic Zeeman energy $g\mu_B SB_{ext}$ of the $Cu^{2+}$ electron spins is comparable to $\Delta$, hence the magnetic field reduces the energy gap as $\Delta(B_{ext}) = \Delta(0) - g\mu_B SB_{ext}$ (*29*), where the *g*-factor of the $Cu^{2+}$ is $g \sim 2.2$ (*30*); we expect $S = 1/2$ for spinons, while $S = 1$ for spin-triplets or two bound-spinons. As shown in Fig.4C, the peak frequency of the NMR lineshape plotted as a function of the normalized frequency $f/f_o - 1$ (= $^{17}K^{(a*)}$) shows a systematic dependence on $B_{ext}$ at low temperatures, implying that $\Delta$, and hence $^{17}K^{(a*)}$, indeed depends on $B_{ext}$. We compare $^{17}K^{(a*)}$ observed for various $B_{ext}$ in Fig.4B and summarize the resulting fitted values of $\Delta(B_{ext})$ in Fig.4D. From the linear fits in Fig.4D, we estimate the zero-field gap as $\Delta(0) \sim 10(3)$ K. Large uncertainties in the slope of the linear fits prevented us from discerning the total spin of the excited state, $S = 1/2$ or $S = 1$. Our estimation of $\Delta(0)/J = 0.03 \sim 0.07$ is comparable to the theoretical prediction based on the DMRG calculations $\Delta(0)/J = 0.07 \sim 0.12$ (*9*). Our results also imply that the Dzyaloshinskii-Moriya interaction is not strong enough to induce appreciable non-zero susceptibility at the lowest measured temperatures. Thus NMR on the single crystal provides strong quantitative evidence for a spin gap in $ZnCu_3(OH)_6Cl_2$. In addition, the ability to distinguish the behavior from different sites near and far from $Cu^{2+}$ defects allows us to conclude that the gapped behavior is intrinsic to the kagome lattice.

**Acknowledgments:**

We thank P. A. Lee and T. Sakai for helpful discussions, and P. Dube, J. Britten, and V. Jarvis of the Brockhouse Institute for technical assistance. The work at McMaster was supported by NSERC and CIFAR. The work at MIT was supported by the US Department of Energy, Office of Science, Office of Basic Energy Sciences under grant no. DE-FG02-07ER46134. T.-H. H. thanks the support from Grainger Fellowship provided by the Department of Physics, University of Chicago

Materials and Methods

Figures S1-S8

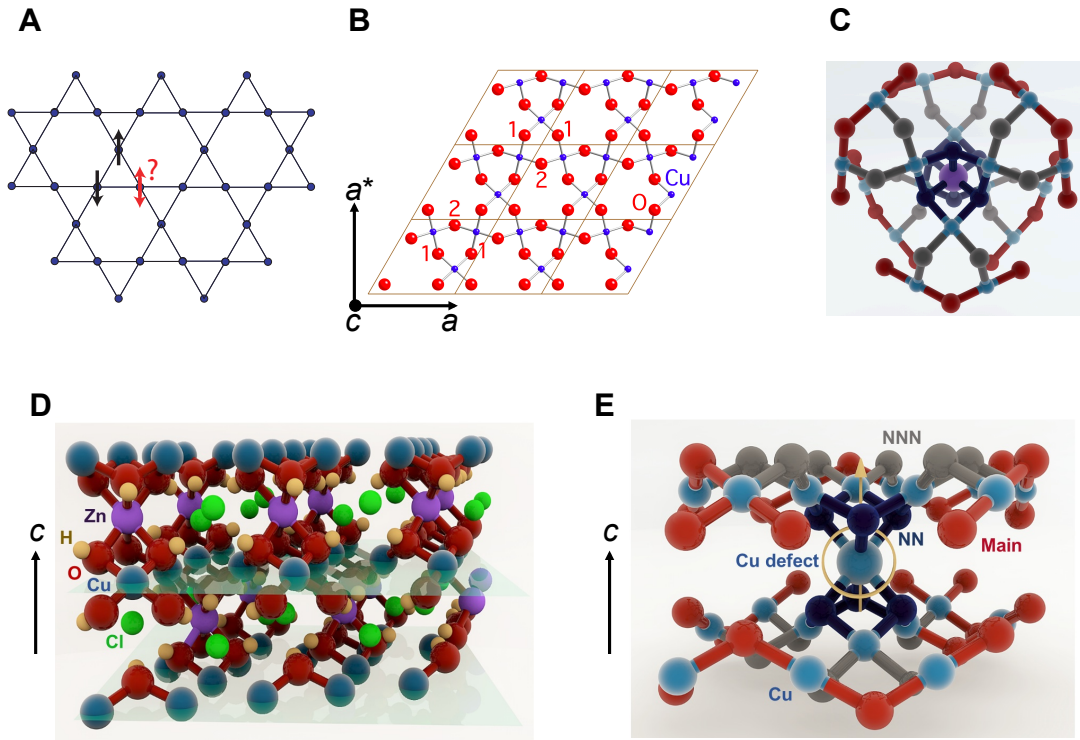

**Fig. 1**. **Kagome lattice and the structure of ZnCu$_3$(OH)$_6$Cl$_2$.** **(A)** A kagome lattice formed by corner-shared triangles. The antiferromagnetic interactions between nearest-neighbor sites frustrate the spin system. **(B)** Top view of the Cu$_3$O$_3$ kagome layer in ZnCu$_3$(OH)$_6$Cl$_2$. The local geometries of the Main1 and Main2 $^{17}$O sites (marked as 1 and 2, respectively) are not equivalent under the presence of $\mathbf{B}_{ext} \parallel$ a*. **(C)** Zn$^{2+}$ sites (purple) are located either above or below the center of Cu$_3$O$_3$ triangles, and have six NN (navy blue) and twelve NNN (grey) $^{17}$O sites in the adjacent kagome planes. **(D)** Oblique view of ZnCu$_3$(OH)$_6$Cl$_2$. **(E)** Cu$^{2+}$ defect moment at a Zn$^{2+}$ site polarized by $\mathbf{B}_{ext} \parallel$ c (gold arrow), and its local environment.

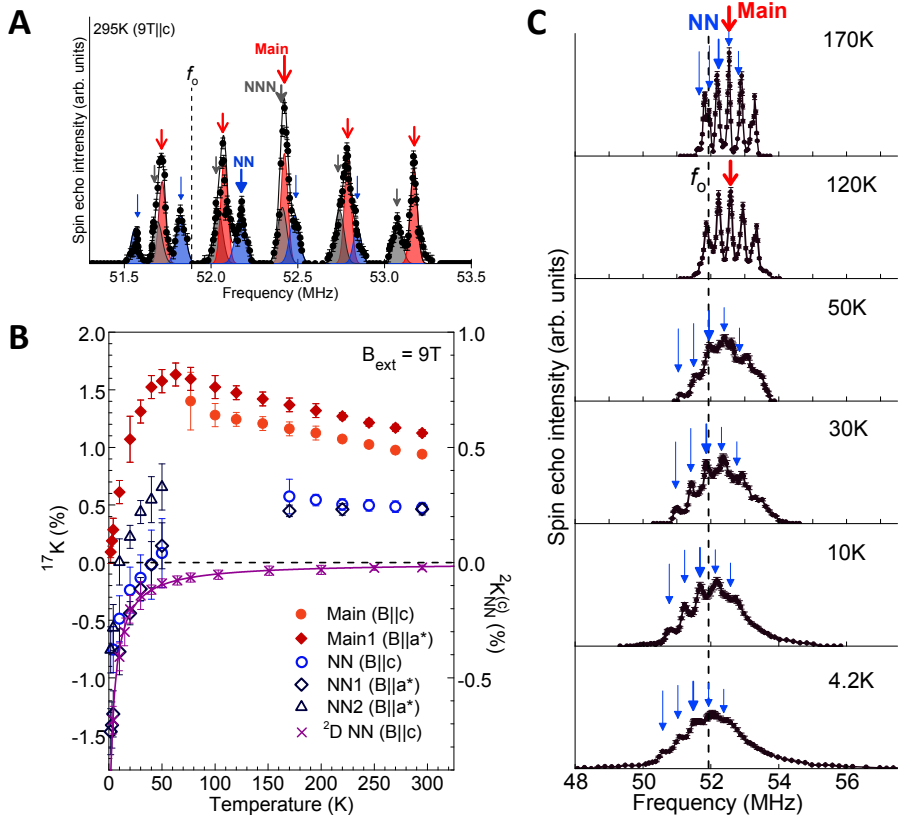

**Fig. 2. Influence of the $Cu^{2+}$ defects and NMR Knight shift in 9 T.** (**A**) $^{17}O$ NMR lineshape at 295 K in $\mathbf{B}_{ext} = 9T \parallel c$, fitted with 3 sets of 5 peaks arising from the Main ($\nu_Q^{(c)} \sim 360$ kHz), NN ($\nu_Q^{(c)} \sim 310$ kHz), and NNN ($\nu_Q^{(c)} \sim 350$ kHz) sites. The vertical dashed line represents the zero-shift frequency $f_o = (\gamma_n/2\pi)B_{ext} \sim 51.9$ MHz, where $^{17}K^{(c)} = 0$. (**B**) A summary of $^{17}K^{(a^*)}$ and $^{17}K^{(c)}$ measured in 9 T. Filled and open symbols represent the data for the Main and NN sites, respectively. Also plotted is the $^2D$ NMR frequency shift $^2K_{NN}^{(c)}$ at the NN $^2D$ sites measured in 8.4 T that is dominated entirely by $\chi_{defect}$; the solid curve is a Curie-Weiss fit, $^2K_{NN}^{(c)} \sim -(T-\theta)^{-1}$, with $\theta \sim -1.2$ K (*23*). (**C**) Representative $^{17}O$ NMR lineshapes measured in $\mathbf{B}_{ext} = 9$ T $\parallel c$ below 295 K.

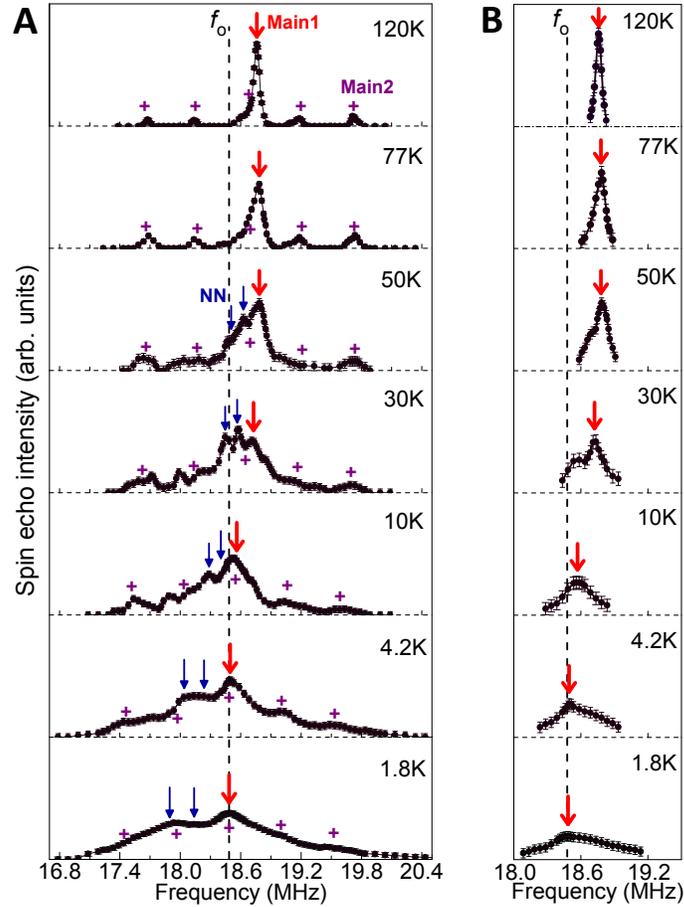

**Fig. 3. $^{17}$O NMR lineshapes measured in $B_{ext} = 3.2$ T || a*.** **(A)** $^{17}$O NMR lineshapes measured with R.F. pulse conditions optimized for individual $I_z = m$ to $m+1$ transitions. Red arrows mark a bundle of all five transitions for Main1. Purple crosses: approximate frequencies of the five individual transitions of Main2; blue arrows: approximate frequencies of the central transitions of the NN1 (left arrows) and NN2 (right arrows) sites (also see Fig.S3). **(B)** The $^{17}$O NMR lineshapes measured with longer R.F. pulses optimized to excite the five $I_z = m$ to $m+1$ transitions of Main1 all at once. Over-pulsing suppresses other peaks. Red arrows in (B) mark the same Main1 peak frequencies as those in the corresponding panels of (A). The dashed vertical line in both panels represents the zero shift frequency $f_o = (\gamma_n / 2\pi) B_{ext} \sim 18.5$ MHz.

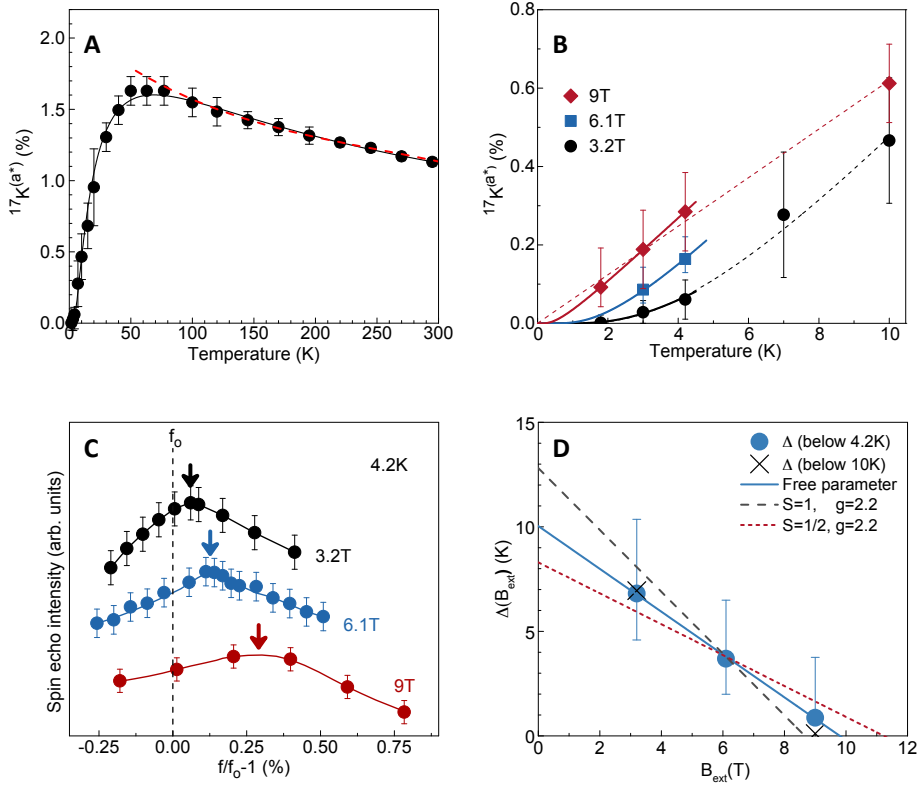

**Fig. 4. Intrinsic spin susceptibility $\chi_{kagome}$ and spin excitation gap $\Delta$.** **(A)** Temperature dependence of $\chi_{kagome}$ deduced from $^{17}K^{(a*)}$ observed at Main1 in $\mathbf{B}_{ext} = 3.2$ T $\parallel$ a*. The red dashed curve represents a theoretical prediction based on high temperature series expansion (*28*) with $J = 180$ K, matched at 295 K, whereas the solid curve is a guide to the eye. **(B)** Temperature and field dependences of $^{17}K^{(a*)}$ at low temperatures, with a fit to $^{17}K^{(a*)} \sim T \cdot \exp(-\Delta/T)$ in the temperature range up to 4.2 K (solid curves) and 10 K (dashed curves). **(C)** Main1 lineshapes at 4.2 K in $\mathbf{B}_{ext} = 3.2, 6.1,$ and 9 T $\parallel$ a* plotted as a function of the normalized frequency $f/f_o - 1$ ($= {}^{17}K^{(a*)}$). **(D)** The spin excitation gap, $\Delta(B_{ext})$, deduced from (B) for the fitting range up to 4.2 K (filled circles) and 10 K (crosses). Dashed and doted lines are the best fits under the constraint of $S = 1$ and $S = 1/2$, respectively, whereas the solid line represents the best free parameter fit.

**Supplementary Materials**

**1. Materials and Methods:**

We grew the single crystal sample of $ZnCu_3(OH)_6Cl_2$ based on the same method as described in detail in (*8*). We enriched the crystal with $^{17}O$ isotope using enriched-water ($H_2^{17}O$) as a starting ingredient. Approximately ~10% of the $^{16}O$ sites were replaced with the $^{17}O$ isotope. We developed a compact homemade two-axis goniometer that could be used with both the X-ray diffractometer and cryogenic NMR probes. We glued our crystal (total mass ~ 40 mg) to the goniometer with an epoxy, and aligned it along the a*- or c-axis orientation by X-ray diffraction. Subsequently, we attached the goniometer to our NMR probe without changing the sample geometry. This procedure enabled us to achieve good alignment for the NMR measurements, despite the highly irregular shape of the crystal. We conducted all NMR measurements using standard pulsed NMR spectrometers. We used a pulse separation time $\tau$ = 15 μs for all the NMR lineshape measurements throughout this work. We measured the NMR lineshapes in Fig.3B using longer R.F. pulses optimized to excite the unsplit NMR peak of main1 consisting of all the five $I_z = m$ to $m+1$ transitions. Since such R.F. pulses are broader by a factor of $\sqrt{5}$ to $\sqrt{9}$ than those optimized for the individual $I_z = m$ to $m+1$ transitions, over-pulsing suppresses the spin echo signals from the main2 and NN sites. We measured the transverse relaxation time $T_2$ by varying the pulse separation time $\tau$, whereas the spin-lattice relaxation time $T_1$ was measured by varying the delay time after an inversion pulse.

## 2. Supplementary Figures

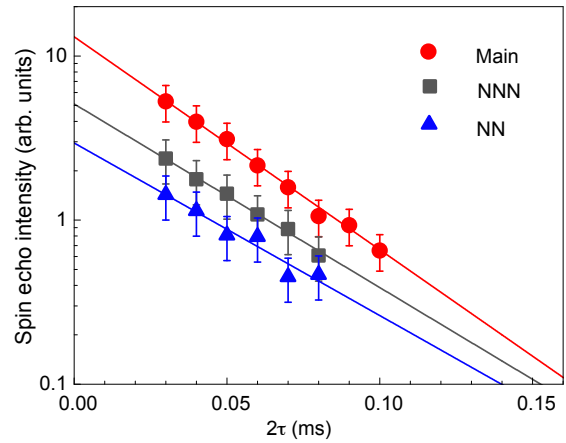

**Fig.S1. Transverse spin echo decay.**

Examples of the transverse ($T_2$) spin echo decay measured at 295 K in $B_{ext}$ = 9 $T \parallel c$ for the uppermost satellite peak of the main and NNN sites, and the lowermost satellite peak of the NN sites. We took into account the effect of the different spin echo decay rate on the integrated intensity of the peaks in Fig.2A in our estimation of their relative intensities.

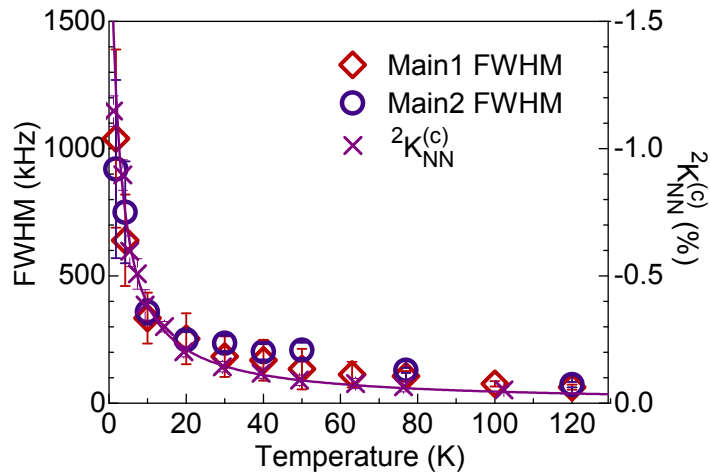

**Fig.S2. Magnetic NMR line broadening caused by defects.**

The Full Width at Half Maximum (FWHM) of the main1 peak and the uppermost satellite-peak of main2 measured in $B_{ext}$ = 3.2T $\parallel a^*$, in comparison to $^2K_{NN}^{(c)}$. The solid curve is a Curie-Weiss fit, $^2K_{NN}^{(c)} \sim -(T-\theta)^{-1}$, with $\theta \sim -1.2$ K (24).

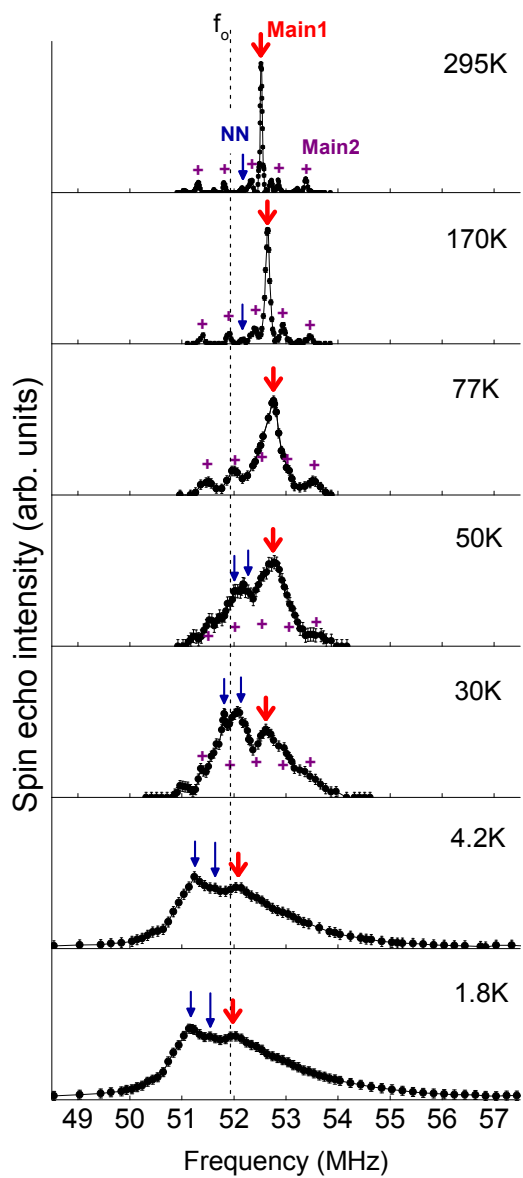

**Fig. S3. Representative $^{17}$O NMR lineshapes measured in $B_{ext} = 9T \parallel a^*$.**
Since the magnetic line broadening caused by $Cu^{2+}$ defects is proportional to the magnitude of the applied magnetic field $B_{ext}$, the 9 T lineshapes have lower resolutions than the 3.2 T lineshapes in Fig. 3A. The purple crosses and blue arrows mark the approximate frequencies of the individual transitions of main2 ($\nu_Q^{(a^*)} \sim 520$ kHz) and the central transitions of the NN1 and NN2 sites, respectively. We tentatively assign the NN1 (left) and NN2 (right) peaks based on their relative intensities. We used the Knight shift of the NN sites estimated in Fig.S3 to plot the corresponding blue arrows in Fig. 3A.

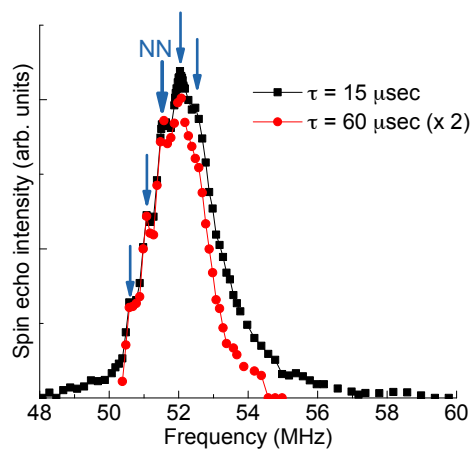

**Fig. S4. Origin of the asymmetric line broadening at low temperatures.**
The magnetic line broadening of the NMR lineshapes observed in $B_{ext}$ = 9 T becomes noticeably asymmetric at low temperature in Fig. 2C and Fig. S3. (The asymmetry is less noticeable in Fig. 3, because the magnetic line broadening is suppressed by a factor of ~3 due to the smaller $B_{ext}$.) Here we compare the lineshapes measured with $\tau$ = 15 μs and 60 μs measured at 4.2 K for $B_{ext}$ = 9 T ∥ c. The broad tail is suppressed for $\tau$ = 60 μs because of the shorter transverse relaxation time $T_2$, and the remaining main peak shows roughly symmetrical line shape. The existence of the high frequency tail with large positive Knight shifts and fast $T_2$ may be an indication that the defect spins induce a short-range spin density oscillation, and some NNN sites have large, positive Knight shifts.

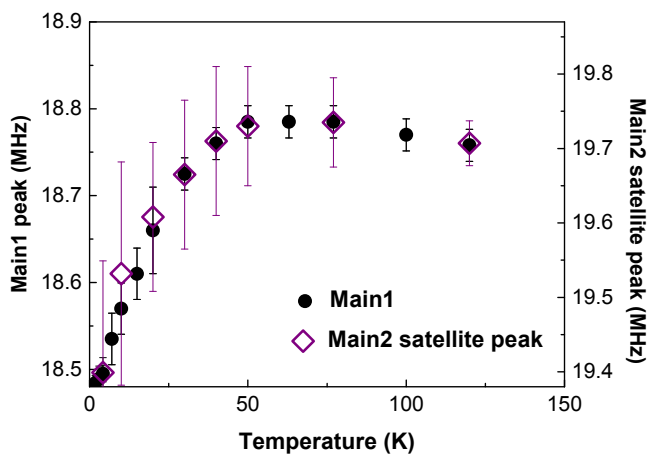

**Fig. S5. Comparison of the peak frequency between main1 and main2.**
We were unable to resolve the central transition of the main2 peak in Fig. 3A due to the superposition by the overwhelmingly strong peak of the main1 sites. Since the peak frequencies of the four additional satellite transitions are affected by the nuclear quadrupole effect, it is difficult to estimate the Knight shift $^{17}K^{(a*)}$ accurately for main2, even though the uppermost satellite peak is isolated and separately observable except below 4.2 K. Nonetheless, we confirmed that the temperature dependence of the peak frequency of the uppermost satellite transition in Fig. 3A shows identical trend as that of main1, as summarized here. This finding implies that $^{17}K^{(a*)}$ exhibits the same trend between main1 and main2.

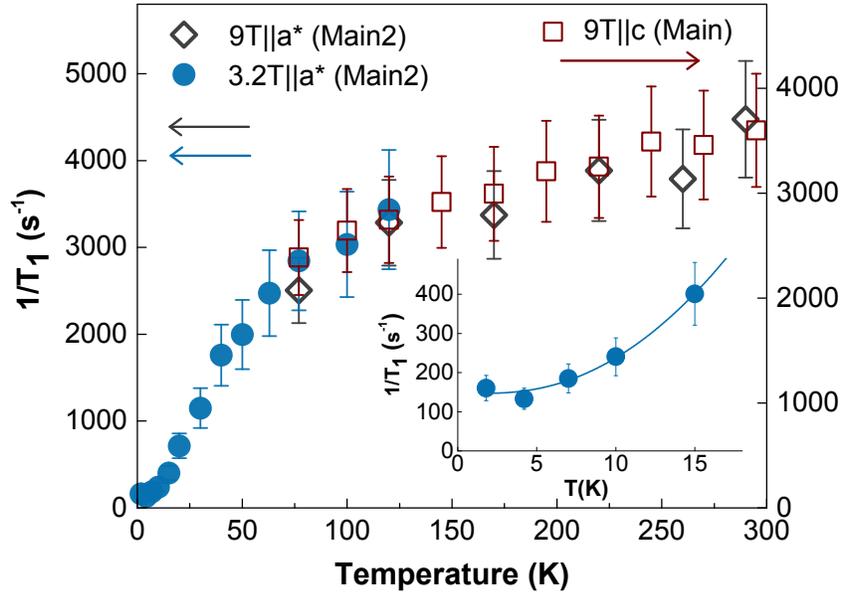

**Fig. S6**: **Low frequency Cu spin dynamics as probed by $1/T_1$.**

We measured the $^{17}$O nuclear spin-lattice relaxation rate $1/T_1$ at the intrinsic main sites using the isolated uppermost satellite transition with $B_{ext} \parallel c$ or $B_{ext} \parallel a*$. For nuclear spin $I = 5/2$, the fundamental transition rate of different normal modes is different by up to a factor of 15; it is therefore essential to use an isolated peak to obtain reliable results of $1/T_1$. $1/T_1 \sim T \Sigma_\mathbf{q} |A(\mathbf{q})|^2 \chi''(\mathbf{q}, f)$ measures the low frequency electron spin dynamics weighted by the hyperfine form factor $|A(\mathbf{q})|^2$; where $\chi''(\mathbf{q}, f)$ is the imaginary part of the dynamical electron spin susceptibility, $\mathbf{q}$ the wave vector within the first Brillouin zone, and $f$ the NMR frequency. The broadening of the peak and/or superposition of other peaks prevented us from measuring $1/T_1$ below ~70 K for both 9T $\parallel$ a* and 9T $\parallel$ c geometries, while the extremely weak signal intensity made $1/T_1$ measurements difficult above ~130 K in 3.2 T $\parallel$ a*. We found that the standard fit of the recovery of the nuclear magnetization for the second satellite transition of $I = 5/2$ is good except below ~5 K, where the residual relaxation rate due to random Cu$^{2+}$ defects at Zn$^{2+}$ sites $(1/T_1)_{defect} \sim 150$ s$^{-1}$ dominates the observed $1/T_1$. The steep decrease of $1/T_1$ observed below

~60 K is qualitatively consistent with the gap signature observed for $^{17}K^{(a^*)}$. In the inset, we show a fit of the temperature dependence below ~15 K with an empirical formula, $1/T_1 = (1/T_1)_{defect} + A\,T^2 \exp(-\Delta/T)$, where we assumed that $\chi''(\mathbf{q}, f) \sim T \cdot \exp(-\Delta/T)$ obeys the same gapped behavior as $^{17}K^{(a^*)}$. We obtained $\Delta \sim 7$ K in 3.2 T $\parallel$ a* from the fit of $1/T_1$, in good agreement with the result from $^{17}K^{(a^*)}$.

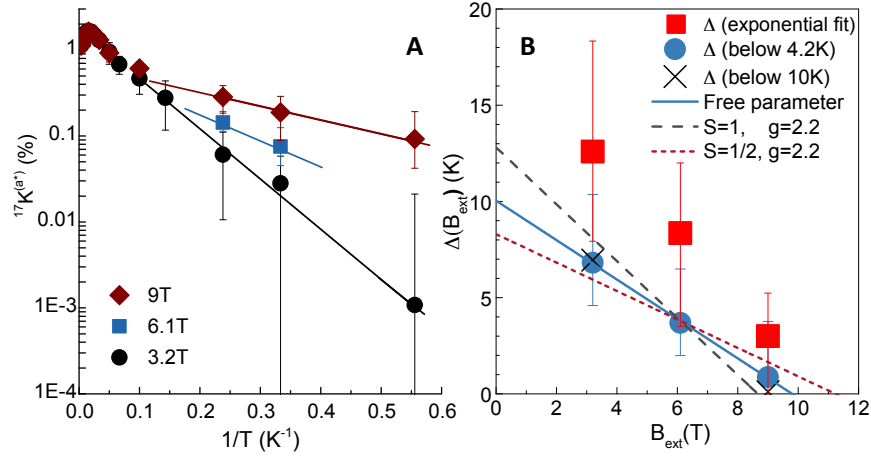

**Fig. S7. Alternate fitting method for the temperature dependence of $^{17}K^{(a^*)}$.**

(A) Solid lines represent an alternate fit of $^{17}K^{(a^*)}$ based on an exponential function, $^{17}K^{(a^*)} \sim \exp(-\Delta/T)$, without the linear pre-factor $T$ used for Fig. 4B. (B) Red squares represent $\Delta$ estimated from the exponential fit in (A), in comparison with the results in Fig. 4D based on the fit with $^{17}K^{(a^*)} \sim T * \exp(-\Delta/T)$. Note that the exponential fit ignores the effect of antiferromagnetic short-range order and attributes the decrease of $^{17}K^{(a^*)}$ entirely to the gap. Accordingly, the results of the exponential fit presented here should be considered the upper-bound of $\Delta$. Nonetheless, the magnitude of $\Delta$ is comparable to the results in Fig.4D within a factor of ~2, and more importantly, the field dependence exhibits qualitatively the same behavior.

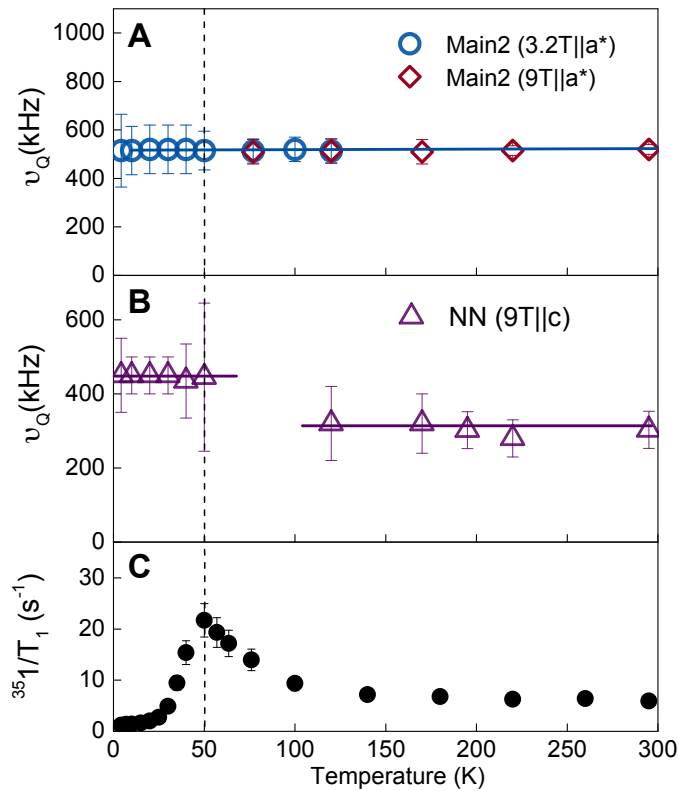

**Fig. S8. Absence of structural distortion at the main $^{17}$O sites.**

The temperature dependence of $\nu_Q$ at (A) the main2 and (B) NN $^{17}$O sites. We observed no change of $\nu_Q$ at the main site, in agreement with the earlier observation based on X-ray diffraction that the average crystal structure does not change as a function of temperature (*15*). On the other hand, $\nu_Q$ at the NN sites below ~50 K is noticeably different from the high temperature value above ~100 K. The line superposition by the main NMR signals prevented us from estimating $\nu_Q$ at the NN sites between ~50 K and ~100 K by using any geometries or field values. Earlier powder $^{35}$Cl NMR measurements also detected enhancement of the $^{35}$Cl nuclear spin-lattice relaxation rate $^{35}1/T_1$ below ~100 K (*22*), as confirmed here for $B_{ext}$ = 3.2 T || a* in (C). The observed change of $\nu_Q$ in (B), combined with the $^{35}1/T_1$ data in (C), suggests that the lattice near the Cu$^{2+}$ defects gradually develops a mild distortion below ~100 K.

## 3. Comparison with earlier powder $^{17}$O NMR measurements

The site-selective anomalous X-ray scattering experiments reported in (*15*) demonstrated that $Cu^{2+}$ defect spins occupy the $Zn^{2+}$ sites outside the kagome planes with ~15% probability, and there are no other defects in herbertsmithite; the upper bound of the population of the anti-site $Zn^{2+}$ defects occupying the $Cu^{2+}$ sites within the kagome plane is estimated to be as little as 1% (*15*). That is, the actual composition of herbertsmithite is $(Zn_{1-x}Cu_x)Cu_3(OH)_6Cl_2$ with x = 0.15 (*i.e.* 15%). Our present single crystal $^{17}$O NMR measurements showed no evidence for the existence of $^{17}$O NMR signals arising from the $Zn^{2+}$ anti-site defects within the kagome planes, in agreement with (*15*). This finding is consistent with our earlier $^2$D single crystal NMR results (*24*); both $^{17}$O and $^2$D NMR data exhibit a split-off peak with ~15% relative intensity. The observed relative intensity agrees with the population of the $Cu^{2+}$ defect spins at the $Zn^{2+}$ sites, and hence allows us to assign them as the NN $^{17}$O and $^2$D sites. Moreover, the temperature dependence of the Knight shift at these $^{17}$O and $^2$D sites obey the same Curie-Weiss behavior at low temperatures as the bulk susceptibility data, providing additional evidence that these split-off NMR peaks arise from the NN of the $Cu^{2+}$ defects occupying the $Zn^{2+}$ sites.

In contrast, an earlier report based on powder Rietveld refinement and powder $^{17}$O NMR measurements concluded that as many as ~5% of the $Cu^{2+}$ sites within the kagome layer are replaced by non-magnetic $Zn^{2+}$ anti-site defects (*23*). Since such a conclusion would strongly influence the interpretation of all the experimental data on herbertsmithite over the last decade, it would be useful to identify the source of discrepancy. First and foremost, it is important to note that the Rietveld refinement reported in (*23*) was conducted under an artificial constraint that the chemical composition is $(Zn_{1-x}Cu_x)(Cu_{1-x/3}Zn_{x/3})_3(OH)_6Cl_2$, by fixing the ratio of the total number of Zn and Cu ions as 1 vs. 3. Such a constraint is equivalent as making an *assumption* that x/3 % of $Zn^{2+}$ anti-site defects exist within the kagome plane when $Cu^{2+}$ defect spins occupy x % of the $Zn^{2+}$ sites. Since x ~ 15 % in typical herbertsmithite samples, the authors of (*23*) inevitably concluded that the population of anti-site defects is x/3 = 15/3 = 5%. We emphasize that their finding of the 5% anti-site defect population is merely the consequence of an assumption, and does not prove the presence of the non-magnetic $Zn^{2+}$ ions within the kagome plane. In fact, the aforementioned anomalous X-ray scattering experiments as well as Rietveld refinement on powder neutron diffraction data analyzed without this constraint in (*15*) established that the artificial constraint used in (*23*) is not justified.

Due to the inherently low resolutions of the NMR lines in powder experiments, the authors of (*23*) were able to detect only one set of defect-induced split-off $^{17}$O NMR signals dubbed as "D-sites," and assigned them to the non-existing $^{17}$O sites adjacent to the anti-site defects. Unlike our single crystal NMR measurements conducted with higher resolutions, they were unable to clearly separate the peak of the "D-sites" at low temperatures, either, and hence concluded that the NMR Knight shift $K^{(D)}$ at the "D-sites" levels off at a *positive* value. Such alleged leveling of $K^{(D)}$ was attributed to the formation of singlet spins at two kagome $Cu^{2+}$ sites adjacent to the anti-site defects. Comparison of the powder $^{17}O$ NMR results of the "D-sites" with our single crystal data reveals, however, that the authors of (*23*) misidentified the NN $^{17}O$ sites of $Cu^{2+}$ defects as the singlet "D-sites." First, we point out that the relative intensity of the "D sites" is ~15% of the overall intensity. Second, the magnitude of the Knight shift $K^{(D)}$ observed for the "D-sites" near room temperature is the same as our NN sites. Third, the field-swept $^{17}O$ powder NMR line of the "D-sites" in Fig.1 of (*23*) (the region marked by blue) appears on the higher field side of the zero shift position at low temperatures; this means that $K^{(D)}$ actually becomes *negative* at low temperatures, even though the summary plot of $K^{(D)}$ in their Fig. 2 somehow presents *positive* values of $K^{(D)}$ at the base temperature. The negative values of the Knight shift observed for their "D-sites" is actually consistent with our observation for the NN $^{17}O$ sites at low temperatures in our Fig. 2B.